\documentclass[doublecol]{epl2}
\usepackage{amssymb,amsmath,graphicx,setspace,color,rotating,subfigure,url}
\usepackage{extarrows}
\usepackage{CJK}
\title{Triadic time series motifs}

\shorttitle{Triadic time series motifs} 

\author{Wen-Jie Xie\inst{1,2} \and Rui-Qi Han\inst{3} and Wei-Xing Zhou\inst{1,2,3}\footnote{e-mail: wxzhou@ecust.edu.cn}}
\shortauthor{W.-J. Xie \etal}

\institute{
  \inst{1} Department of Finance, East China University of Science and Technology, Shanghai 200237, China\\
  \inst{2} Research Center for Econophysics, East China University of Science and Technology, Shanghai 200237, China\\
  \inst{3} Department of Mathematics, East China University of Science and Technology, Shanghai 200237, China\\
}

\abstract{
  We introduce the concept of time series motifs for time series analysis. Time series motifs consider not only the spatial information of mutual visibility but also the temporal information of relative magnitude between the data points. We study the profiles of the six triadic time series. The six motif occurrence frequencies are derived for uncorrelated time series, which are approximately linear functions of the length of the time series. The corresponding motif profile thus converges to a constant vector $(0.2,0.2,0.1,0.2,0.1,0.2)$. These analytical results have been verified by numerical simulations. For fractional Gaussian noises, numerical simulations unveil the nonlinear dependence of motif occurrence frequencies on the Hurst exponent. Applications of the time series motif analysis uncover that the motif occurrence frequency distributions are able to capture the different dynamics in the heartbeat rates of healthy subjects, congestive heart failure (CHF) subjects, and atrial fibrillation (AF) subjects and in the price fluctuations of bullish and bearish markets. Our method shows its potential power to classify different types of time series and test the time irreversibility of time series.
}

\begin{document}
\maketitle



\section{Introduction}
\label{S1:Introduction}

Time series analysis has been widely used in diverse fields~\cite{Brockwell-Davis-1991,Mandelbrot-1970-Em,
Dickey-Fuller-1981-Em,Bouchaud-Muzy-2003-LNP,Bouchaud-Potters-Meyer-2000-EPJB,Krawiecki-Holyst-Helbing-2002-PRL,Li-Yang-Komatsuzak-2008-PNAS,Peng-Havlin-Stanley-Goldberger-1995-Chaos}. In recent years, there is a huge increase of interest in application of complex network theory
\cite{Albert-Barabasi-2002-RMP,Dorogovtsev-Mendes-2002-AdvPhys,Newman-2003-SIAMR,Boccaletti-Latora-Moreno-Chavez-Hwang-2006-PR,Costa-Oliveira-Travieso-Rodrigues-Boas-Antiqueira-Viana-Rocha-2011-AdvPhys, Barthelemy-2011-PR,Holme-Saramaki-2012-PR,Boccaletti-Bianconi-Criado-delGenio-GomezGardenes-Romance-SendinaNadal-Wang-Zanin-2014-PR}
to analyze time series \cite{Gao-Small-Kurths-2016-EPL}, such as cycle networks \cite{Zhang-Small-2006-PRL,Zhang-Sun-Luo-Zhang-Nakamura-Small-2008-PD}, nearest neighbor networks \cite{Xu-Zhang-Small-2008-PNAS}, $n$-tuple networks \cite{Li-Wang-2006-CSB,Li-Wang-2007-PA}, recurrence networks \cite{Zou-Pazo-Romano-Thiel-Kurths-2007-PRE,Marwan-Romano-Thiel-Kurths-2007-PR,Marwan-2008-EPJST,Donner-Zou-Donges-Marwan-Kurths-2010-NJP,Marwan-Donges-Zou-Donner-Kurths-2009-PLA}, visibility graphs (VGs) \cite{Lacasa-Luque-Ballesteros-Luque-Nuno-2008-PNAS,Luque-Lacasa-Ballesteros-Luque-2009-PRE,Gao-Cai-Yang-Dang-Zhang-2016-SR,Gao-Cai-Yang-Dang-2017-PA,Bianchi-Livi-Alippi-Jenssen-2017-SR}, the segment correlation network \cite{Yang-Yang-2008-PA}, and temporal graphs \cite{Shirazi-Jafari-Davoudi-Peinke-Tabar-Sahimi-2009-JSM,Kostakos-2009-PA}. The methods of visibility graphs and horizontal visibility graphs (HVGs) \cite{Lacasa-Toral-2010-PRE} have been widely applied in financial markets \cite{Ni-Jiang-Zhou-2009-PLA,Qian-Jiang-Zhou-2010-JPA,Yang-Wang-Yang-Mang-2009-PA,Vamvakaris-Pantelous-Zuev-2018-PA,Li-Zhao-2018-EPL}, biological systems \cite{Lacasa-Luque-Luque-Nuno-2009-EPL,Shao-2010-APL,Dong-Li-2010-APL,Ahmadlou-Adeli-Adeli-2010-JNT}, ecological systems \cite{Elsner-Jagger-Fogarty-2009-GRL,Tang-Liu-Liu-2010-MPLB}, and some other complex systems \cite{Liu-Zhou-Yuan-2010-PA,Xie-Zhou-2011-PA,Ahadpour-Sadra-2012-IS,Fan-Guo-Zha-2012-PA,Lacasa-2014-NL,Lacasa-2016-JPA,Xie-Han-Jiang-Wei-Zhou-2017-EPL}, to list a few.

Network motifs are usually regarded as the building blocks of complex networks \cite{Milo-Itzkovitz-Kashtan-Levitt-ShenOrr-Ayzenshtat-Sheffer-Alon-2004-Science,Milo-ShenOrr-Itzkovitz-Kashtan-Chklovskii-Alon-2002-Science,Milo-Kashtan-Itzkovitz-Newman-Alon-2004-XXX}. Their occurrence patterns are used to define superfamilies of networks \cite{Milo-ShenOrr-Itzkovitz-Kashtan-Chklovskii-Alon-2002-Science,Milo-Itzkovitz-Kashtan-Levitt-ShenOrr-Ayzenshtat-Sheffer-Alon-2004-Science}, which is able to reflect the evolution dynamics of complex systems \cite{Kovanen-Kaski-Kertesz-Saramaki-2013-PNAS,Klimek-Thurner-2013-NJP}. Based on the structural similarity between motif distributions in nearest neighbor networks mapped from time series, the different types of dynamics in periodic, chaotic and noisy processes can be distinguished \cite{Xu-Zhang-Small-2008-PNAS}. Moreover, the networks mapped from time series with different dynamics of periodic, chaotic and noisy processes are found to exhibit different motif distributions \cite{Xu-Zhang-Small-2008-PNAS}.

Recently, Iacovacci and Lacasa introduced the sequential HVG motifs \cite{Iacovacci-Lacasa-2016a-PRE} and the sequential VG motifs \cite{Iacovacci-Lacasa-2016b-PRE}. The sequential HVG motifs of $s$ nodes are defined as the HVG subgraphs formed in $s$ successive data points of the time series and there are two distinct triadic sequential HVG motifs and six distinct tetradic sequential HVG motifs \cite{Iacovacci-Lacasa-2016a-PRE}.
The sequential VG motifs of size $s$ are defined as the VG subgraphs formed in $s$ successive data points of the time series and there are eight distinct tetradic sequential HVG motifs \cite{Iacovacci-Lacasa-2016b-PRE}.

In this Letter, we propose a new concept of time series motifs for time series analysis. Time series motifs of size $s$ are determined by the relative magnitude among $s$ data points that are randomly chosen from the time series. We focus on triadic motif profiles. The analytic profile of triadic time series motifs can be derived for uncorrelated time series, which is consistent with numerical results. For correlated time series, we determine the triadic motif profiles numerically with fractional Brownian motions. We further study two real time series in physiology and finance.

\section{Constructing triadic time series motifs}

\begin{figure}[t!]
  \centering
  \includegraphics[width=8cm]{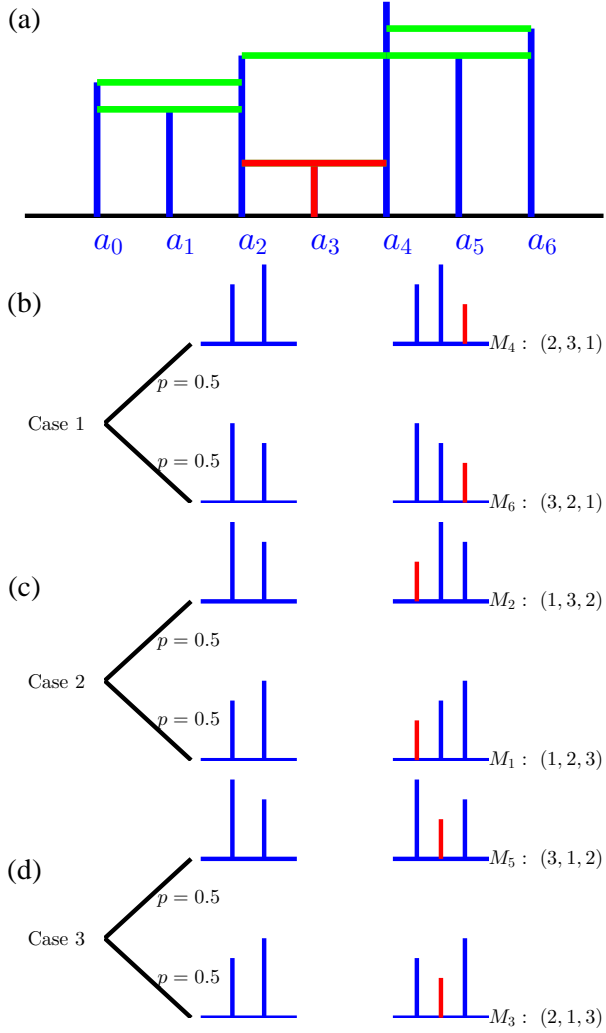}
  \caption{\label{Fig:TSM:Triadic:Def} (Color online.) Definition of triadic time series motifs and illustration of the iterative construction process.}
\end{figure}

Let us start from diadic time series motifs. For two arbitrary points $x_i$ and $x_j$ $(i<j)$, a motif forms if and only if the data points between $x_i$ and $x_j$ are less than $x_i$ and $x_j$:
\begin{equation}
  x_i > x_n ~~~{\mathrm{and}}~~~ x_j > x_n,
\end{equation}
where $i<n<j$. As shown in the left plot of fig.~\ref{Fig:TSM:Triadic:Def}B, there are two admissible motifs according to their relative magnitudes:
\begin{equation}
  x_i > x_j ~~~{\mathrm{or}}~~~ x_i \leq x_j.
\end{equation}
For simplicity, we can denote respectively these two motifs as $(1,2)$ and $(2,1)$. A monotonically increasing time series such as the Devil's Stair has only one type of motifs, that is $(1,2)$, and vice versa.

Triadic time series motifs can form from three arbitrary data $x_i$, $x_j$ and $x_k$ with $i<j<k$ that satisfy the following conditions:
\begin{equation}
   \left\{
    \begin{array}{lllll}
     x_i>x_n &{\rm{and}}& x_j>x_n, &  \forall n\in(i,j) \\
     x_j>x_m &{\rm{and}}& x_k>x_m, &  \forall m\in(j,k)
    \end{array}
    \right..
    \label{Eq:HVG}
\end{equation}
The admissible triadic motifs are shown in fig.~\ref{Fig:TSM:Triadic:Def}, which can be denoted as $M_1=(1,2,3)$, $M_2=(1,3,2)$, $M_3=(2,1,3)$, $M_4=(2,3,1)$, $M_5=(3,1,2)$ and $M_6=(3,2,1)$. Among these motifs, $M_1$, $M_2$, $M_4$ and $M_6$ are open, while $M_3$ and $M_5$ are closed.
Time series motifs of higher size can be defined similarly and their symbolic expressions are the permutations of elements $(1,2,\cdots,s)$. Hence, the number of admissible motifs of size $s$ is $s!$. In this Letter, we focus on triadic time series motifs.

The time series motifs defined in this Letter are different from the sequential HVG motifs \cite{Iacovacci-Lacasa-2016a-PRE} and the convention HVG motifs. If we do not distinguish respectively the four open motifs and the two closed motifs, we obtain the two convention HVG motifs. The concept of time series motifs is different because it considers not only the spatial information (visibility) but also the temporal information (relative magnitude) between the data points under consideration. Nevertheless, for convenience, we can still use the concept of HVG in our analysis. We note that both the sequential HVG motifs and the time series motifs share some common features with the ordinal patterns \cite{Keller-Sinn-2005-PA}.

\section{Motif profiles of uncorrelated time series}

To derive the motif profiles of uncorrelated time series, we adopt the iterative construction process proposed in Ref.~\cite{Xie-Han-Jiang-Wei-Zhou-2017-EPL}. We first generate an uncorrelated time series of size $T$ and preset a one-dimensional lattice of $T$ positions. We then sort the data points in a decreasing order and assign them one by one to $T$ random positions.
In the first step, we randomly choose a position and assign the largest number to it.
In the $t$th step, we randomly choose a position from the rest $(T-t+1)$ unassigned positions and assign the $t$th largest number to it.

As a result, the new node is the smallest among the $t$ numbers and only two edges are added to the HVG. As shown in fig.~\ref{Fig:TSM:Triadic:Def}(a), the red vertical line indicates the new node and the red horizontal lines indicate the two new edges.
We have to note that the $t$th node may be placed at the beginning of or at the end of the $t-1$ nodes. In this case only one edge is added.
However, when $t$ is large enough, we can ignore this end-point effect.
From fig.~\ref{Fig:TSM:Triadic:Def}(a), the two new edges added in the $t$th step don't affect the motifs already existing in the $(t-1)$th step.
Comparing with the $(t-1)$th step, the number of motifs presented in the $t$th step increases.
After adding the $t$th node denoted $a_3$ in fig.~\ref{Fig:TSM:Triadic:Def}(a), we consider three cases as illustrated respectively in fig.~\ref{Fig:TSM:Triadic:Def}(b-d). In all cases, and the new motifs must contain node $a_3$ because motifs without $a_3$ have ready formed before the $t$th step.

\begin{figure*}[t!]
  \centering
  \includegraphics[width=16cm]{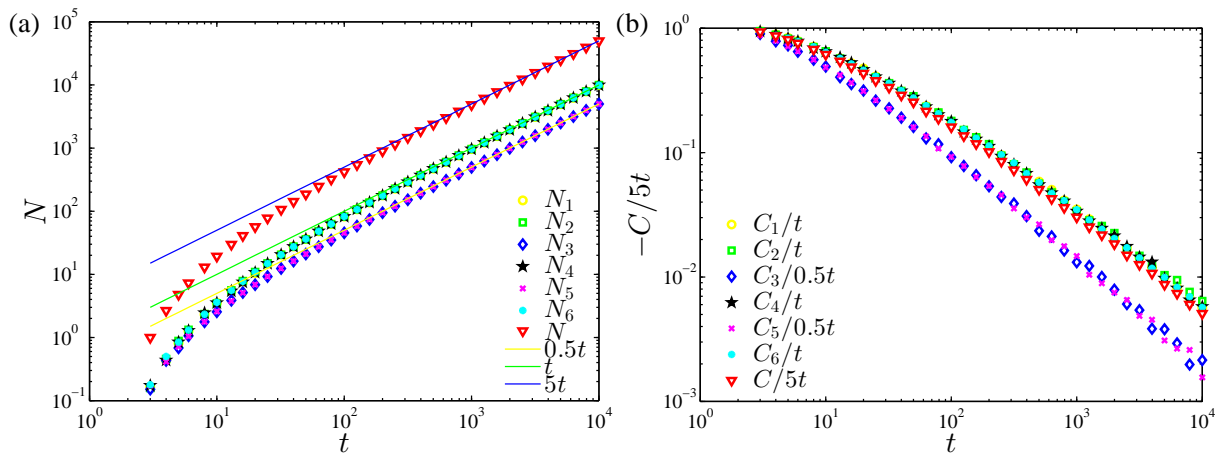}
  \caption{(Color online.) Profiles of triadic time series motifs for uncorrelated time series. We simulated 1000 times for each time series of length $t$. Each data point (markers) represents the average value over the 1000 repeated simulations. The solid lines stands for analytic approximations. (a) Average occurrence numbers of the triadic time series motifs.
  (b) Average relative differences of the motifs.}
  \label{Fig:TSM:Triadic:uncorr:N}
\end{figure*}

As shown in fig.~\ref{Fig:TSM:Triadic:Def}(b), the first case considers node $a_3$ and all the nodes on the left to it. The new motifs in the first case must also contain node $a_{2}$ because $a_{3}$ is only horizontally visible to $a_2$ on the left part, and the third node can be one of the nodes on the left side of node $a_{2}$ which are connected to node $a_{2}$. Therefore, the number of new motifs is the number of nodes that are on the left side of $a_2$ and visible to $a_2$. In other words, the number of new motifs is the in-degree of $a_2$ in the corresponding directed HVG in which a direct link of two horizontally visible data points is formed from the left data point to the right one \cite{Lacasa-Nunez-Roldan-Parrondo-Luque-2012-EPJB,Xie-Han-Jiang-Wei-Zhou-2017-EPL}. Since the in-degree distribution of the associated directed horizontal visibility graph is $P(k_{\rm{in}})=(1/2)^k$  \cite{Lacasa-Nunez-Roldan-Parrondo-Luque-2012-EPJB}, the average in-degree $\langle k_{\rm{in}}\rangle$ is $\sum_{k=1}^{\infty}k(1/2)^k=2$.
It means that the expected number of new motifs introduced in the first case is 2.
As shown in fig.~\ref{Fig:TSM:Triadic:Def}(c), the second case considers node $a_3$ and all the nodes on the right to it.
A similar analysis shows that the number of new motifs in the second case is the average out-degree $\langle k_{\rm{out}}\rangle$ of node $a_{4}$, which equals to 2.
As shown in fig.~\ref{Fig:TSM:Triadic:Def}(d) for the third case, a new triadic motif contains three nodes $a_{2}$, $a_{3}$ and $a_{4}$, which could be $M_3(2,1,3)$ or $M_5(3,1,2)$.

\begin{figure*}[htp]
  \centering
  \includegraphics[width=0.95\linewidth]{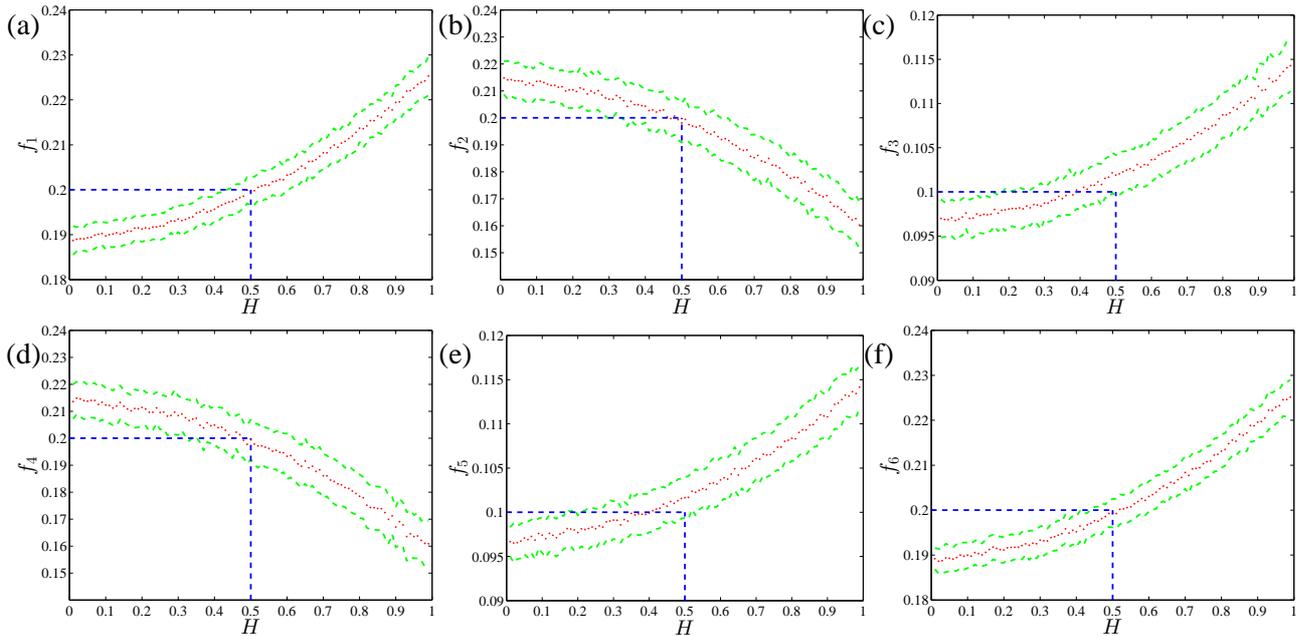}
  \caption{(Color online.) Dependence of the occurrence frequencies $f_i$ of motifs $M_i$ on the Hurst exponents $H$ for fractional Gaussian noises of length $T=1000$. The Hurst exponent $H$ varies from 0.01 to 0.99 with a spacing of 0.01. For each $H$, we simulated 1000 times. The red dots stand for the average occurrence frequencies and the dashed green lines show the corresponding standard deviations. The intersections of dashed blue lines indicate the theoretical values ${\bf{f}}=(0.2,0.2,0.1,0.2,0.1,0.2)$ for $H=0.5$.}
  \label{Fig:TSM:Triadic:fGn}
\end{figure*}

\subsection{Occurrence of all motifs}

\begin{subequations}
On average, 5 new motifs are introduced in each step. Denoting $N(t)$ the number of all motifs at the $t$-th step, we obtain the following iterative equation:
\begin{equation}
  N(t)=N(t-1)+5+\varepsilon(t),
  \label{Eq:NewMotif:sum}
\end{equation}
where $\varepsilon(t)$ is a random variable. The approximate solution to eq.~(\ref{Eq:NewMotif:sum}) is:
\begin{equation}
  N(t)=5t+C(t),
  \label{Eq:NewMotif:Solution:sum}
\end{equation}
\end{subequations}
where $C(t)$ is the sum of differences between the actual number and the theoretical approximate number $5t$ of motifs. We have $N(1)=0$, $N(2)=0$, $N(3)=1$, $N(4)<5$ and the number of new motifs
is less than 5 when $t$ is small. Thus $C(t)<0$. We verify the theoretical analysis through extensive numerical experiments and the results are shown in fig.~\ref{Fig:TSM:Triadic:uncorr:N}.
The number of motifs increases linearly with the length of time series.

\subsection{Occurrence of individual motifs}

In the following, we shall analyze the occurrence frequency of each motif.

In fig.~\ref{Fig:TSM:Triadic:Def}(b), node $a_{1}$ and node $a_{2}$ in the new motif are on the left side of node $a_{3}$ and
their relation satisfies the inequality $a_{1}<a_{2}$ or $a_{1}>a_{2}$.  Because the new node $a_{3}$ is the smallest, we have
$a_{3}<a_{1}$ and $a_{3}<a_{2}$. In the random sequences, we get the new motif $M_4(2,3,1)$ or $M_6(3,2,1)$ with the same
probability $p=0.5$ as shown in fig.~\ref{Fig:TSM:Triadic:Def}(b). Hence, the average number of new motif $M_4(2,3,1)$ or $M_6(3,2,1)$ is $2p=1$.

In fig.~\ref{Fig:TSM:Triadic:Def}(c), node $a_{4}$ and node $a_{5}$ in the new motif are on the right side of node $a_{3}$ and their relation satisfies the inequality $a_{4}<a_{5}$ or $a_{4}>a_{5}$.  Also, we have $a_{3}<a_{4}$ and $a_{3}<a_{5}$. In the random sequences, we get the new motif $M_1(1,2,3)$ or $M_2(1,3,2)$ with the same probability $p=0.5$ as shown in fig.~\ref{Fig:TSM:Triadic:Def}(c).  In this case, the average number of new motifs $M_1(1,2,3)$ or $M_2(1,3,2)$ is $2p=1$.

In fig.~\ref{Fig:TSM:Triadic:Def}(d), node $a_{2}$ and node $a_{4}$ in the new motif are on the both sides of node $a_{3}$ and their relation satisfies the inequality $a_{2}<a_{4}$ or $a_{2}>a_{4}$.  In the same way, we have $a_{3}<a_{2}$ and $a_{3}<a_{4}$ and get the new motif $M_3(2,1,3)$ or $M_5(3,1,2)$ with the same probability $p=0.5$. In this case, we obtain only 1 new motif and the average number of new motifs $M_3(2,1,3)$ or $M_5(3,1,2)$ is $1p=0.5$.

\begin{subequations}
For the open triadic motifs $M_{i}$  ($i\in\{1,2,4,6\}$), the iterative equation of the number of $M_i$ is expressed as follows:
\begin{equation}
      N_{i}(t)=N_{i}(t-1)+1+\varepsilon_{i}(t),~~~i\in\{1,2,4,6\},
    \label{Eq:NewMotif:open}
\end{equation}
where $\varepsilon_{i}(t)$ is a random variable. The approximate solution to eq.~(\ref{Eq:NewMotif:open}) is:
\begin{equation}
  N_{i}(t)=t+C_{i}(t),~~~i\in\{1,2,4,6\},
  \label{Eq:NewMotif:Solution:open}
\end{equation}
\end{subequations}
in which $C_i(t)$ is the sum of differences between the actual number and the theoretical approximate number $t$ of motifs.  We have $N_i(1)=0$, $N_i(2)=0$ and the number of new motifs
is less than 1 when $t$ is small. Thus $C_i(t)<0$. We verify the theoretical analysis through extensive numerical experiments and the results are shown in Fig.~\ref{Fig:TSM:Triadic:uncorr:N}.

\begin{figure*}[t!]
\centering
  \includegraphics[width=0.95\linewidth]{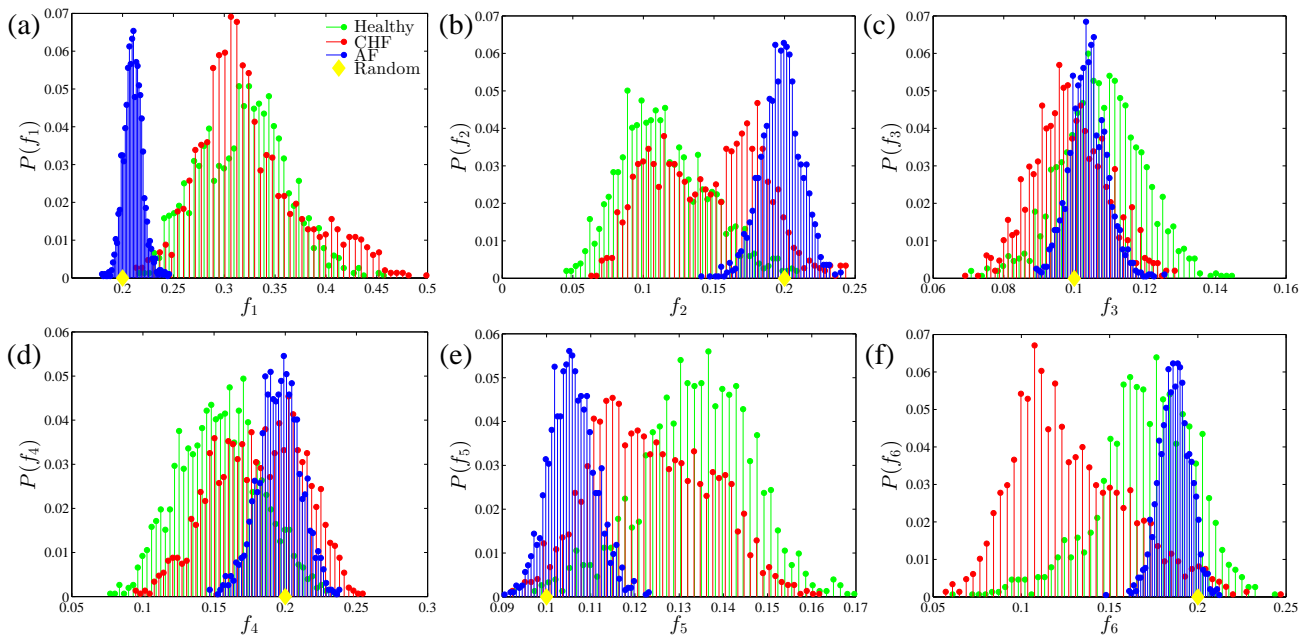}
  \caption{(Color online.) Occurrence frequency distributions $P(f_{i})$ of the six triadic time series motifs identified from five inter-heartbeat interval time series of healthy subjects (green), CHF patients (red) and AF patients(blue). Each time series of length $L$ is partitioned into $ \lfloor L/l \rfloor $ non-overlapping sub-series of length $l=400$. The occurrence frequencies $f_i$ with $i=1,2,\cdots,6$ are obtained for each sub-series. We also draw the six analytical values of uncorrelated time series as filled yellow rhombi for comparison.}
    \label{Fig:TSM:Triadic:RR}
\end{figure*}

\begin{subequations}
For the closed triadic motif $M_{i}$ ($i\in\{3,5\}$), we obtain the following iterative equation for their numbers:
\begin{equation}
  N_{i}(t)=N_{i}(t-1)+0.5+\varepsilon_{i}(t),~~~i\in\{3,5\},
  \label{Eq:NewMotif:closed}
\end{equation}
whose approximate solution is:
\begin{equation}
  N_{i}(t)=0.5t+C_{i}(t),~~~i\in\{3,5\},
  \label{Eq:NewMotif:Solution:closed}
\end{equation}
\end{subequations}
where $C_i(t)$ is the sum of differences between the actual number and the theoretical approximate number $0.5t$ of motifs.  We have $N_i(1)=0$, $N_i(2)=0$ and the number of new motifs
is less than 0.5 when $t$ is small. Thus $C_i(t)<0$. We verify the theoretical analysis through extensive numerical experiments and the results are shown in fig.~\ref{Fig:TSM:Triadic:uncorr:N}.

By definition, we have
$N(t)=\sum_{i=1}^{6}N_{i}(t)$ and $C(t)=\sum_{i=1}^{6}C_{i}(t)$.
Hence, the occurrence frequencies of the six motifs are calculated as
\begin{equation}
  f_i(t) =N_{i}(t)/N(t).
\end{equation}
Combining eq.~(\ref{Eq:NewMotif:Solution:sum}), eq.~(\ref{Eq:NewMotif:Solution:open}) and eq.~(\ref{Eq:NewMotif:Solution:closed}), when $t$ is sufficiently large, the occurrence frequencies of the six motifs are ${\bf{f}}=(0.2,0.2,0.1,0.2,0.1,0.2)$ for uncorrelated time series.

We verify the validity of the theoretical analysis for the occurrence frequencies of the six motifs through extensive numerical experiments. Each data point in fig.~\ref{Fig:TSM:Triadic:uncorr:N} represents the average value over 1000 repeated simulations for each length $t$ of time series. The results shown in fig.~\ref{Fig:TSM:Triadic:uncorr:N}(a) are in very good agreement with the theoretical expected value when $t$ is large enough. In fig.~\ref{Fig:TSM:Triadic:uncorr:N}(b), we show the average relative differences $C/5t$, $C/t$, $C/0.5t$ of the six motifs and the average difference $C<0$. The average absolute values of $C/5t$, $C/t$, $C/0.5t$ decrease with the time series' length $t$ increases.

\section{Motif profiles for fractional Gaussian noises}

Through extensive numerical simulations, we determine the triadic motif profiles extracted from fractional Gaussian noises. The main purpose is to investigate the impacts of correlations in time series as quantified by the Hurst exponent $H$.
The Hurst exponent $H$ varies from 0.01 to 0.99 with a spacing of 0.01. For each Hurst exponent $H$, we simulate 1000 time series of fractional Gaussian noises of length $T=1000$ and determine the occurrence numbers $N_i$ and the occurrence frequencies $f_i$ of the six triadic motifs.

The results are illustrated in fig.~\ref{Fig:TSM:Triadic:fGn}. We find that there are three pairs of motifs whose occurrence frequency curves are identical: $M_1(1,2,3)$ and $M_6(3,2,1)$, $M_2(1,3,2)$ and $M_4(2,3,1)$, and $M_3(2,1,3)$ and $M_5(3,1,2)$. This observation is due to the fact that fractional Gaussian noises are time reversible \cite{Lacasa-Nunez-Roldan-Parrondo-Luque-2012-EPJB}, as illustrated in fig.~\ref{Fig:TSM:Triadic:Def}. If three data points of a given time series $\{x_1, \cdots, x_T\}$ form a motif $M_1$, they form a motif $M_6$ in the reversed time series $\{x_T, \cdots, x_1\}$, which holds for the other two motif pairs as well.

We find that the occurrence frequencies of $M_1$ ($M_6$) and $M_3$ ($M_5$) increase with $H$ and are convex, while the occurrence frequency of $M_2$ ($M_4$) decreases with $H$ and is concave. A comparison of the numerical values and the analytical approximate values of $f_i$ for $H=0.5$ suggests that the approximate analytical analysis under-estimates $f_3$ and $f_5$. With the increase of time series length $T$, the discrepancy narrows.

\section{Applications to real systems}

We apply the concept of triadic time series motifs to two real systems: the human heart inter-beat intervals and the US stock market's DJIA index returns.

\subsection{Human inter-heartbeat intervals}

%

As the first case, we investigate the human inter-heartbeat interval time series of five healthy subjects, five congestive heart failure (CHF) patients, and five subjects suffering from atrial fibrillation (AF). The filtered outlier-free data sets were retrieved from PhysioNet at http://www.physionet.org/challenge/chaos/ \cite{Shao-2010-APL,Dong-Li-2010-APL}. The time series sizes $L$ are respectively $99762$, $86925$, $101523$, $86822$ and $81280$ for the five healthy subjects, are respectively $74496$, $76948$, $88501$, $88499$ and $115062$ for the five CHF patients, and respectively $101145$, $117100$, $85304$, $138209$ and $141628$ for the five AF subjects.

Each time series of length $L$ is partitioned into $ \lfloor L/l \rfloor $ non-overlapping sub-series of length $l=400$. The occurrence frequencies $f_i$ with $i=1,2,\cdots,6$ are obtained for each sub-series. For each of three types of subjects, we put all the $f_i$ values together and calculate the occurrence frequency distribution $P(f_i)$. The $3\times6=18$ resulting distributions are illustrated in fig.~\ref{Fig:TSM:Triadic:RR}, in which we also draw the six analytical values of uncorrelated time series as filled yellow rhombi for comparison.

The first observation is that the occurrence frequency distributions of the AF subjects are around the analytical values of uncorrelated time series. It shows that the heartbeat dynamics of AF subjects are close to be stochastic
\cite{AlvarezRamirez-Rodriguez-Echeverria-2009-Chaos,Aronis-Berger-Calkins-Chrispin-Marine-Spragg-Tao-Tandri-Ashikaga-2018-Chaos}. In contrast, except for $f_3$ of the CHF subjects, the occurrence frequency distributions for healthy and CHF subjects deviate remarkably from the uncorrelated case. In addition, the two distributions for healthy and CHF subjects are different from each other in each plot. It suggests that the heartbeat dynamics of healthy and CHF subjects have very different features, which can be well identified by the occurrence frequencies of triadic time series motifs.

\begin{figure*}[t!]
\centering
  \includegraphics[width=0.95\linewidth]{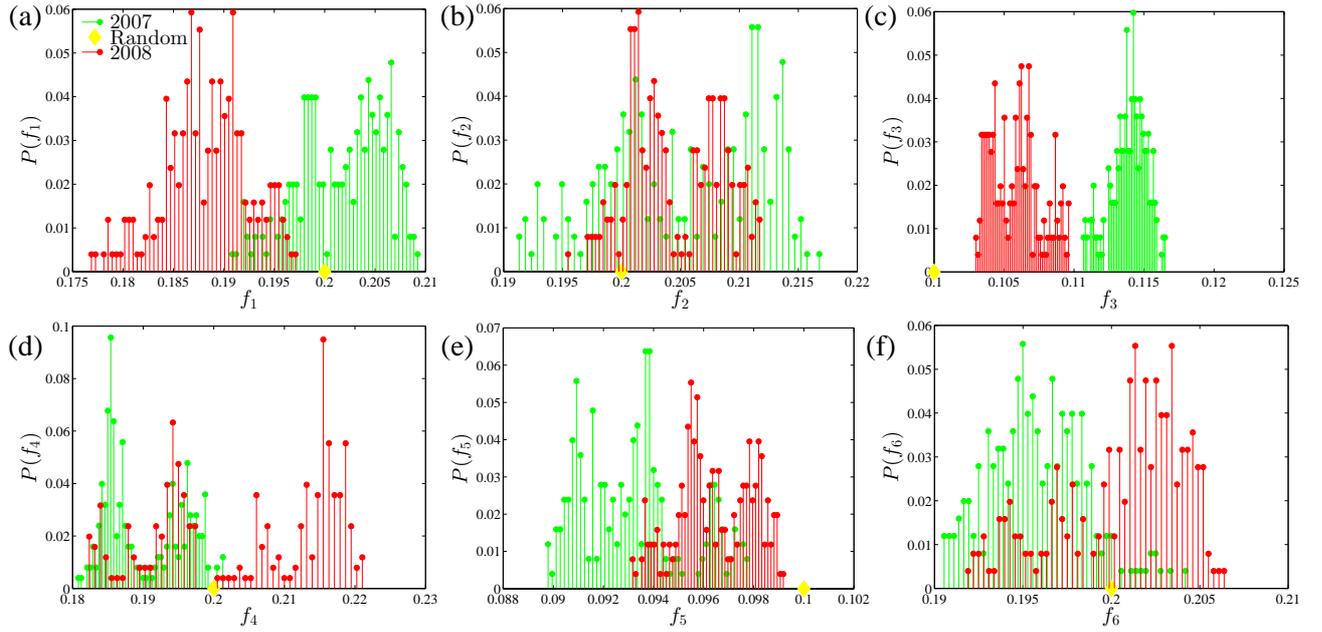}
  \caption{(Color online.) Occurrence frequency distributions $P(f_{i})$ of the six triadic time series motifs identified from the DJIA index returns before and after 31 December 2007. The length of moving windows is 400 trading days. The six filled yellow rhombi indicate the analytical values of uncorrelated time series for comparison.}
    \label{Fig:TSM:Triadic:DJIA}
\end{figure*}

\subsection{DJIA index returns}

We also study the daily DJIA index returns from 2 June 2005 to 3 August 2010. The return time series are divided into two parts delimited on 31 December 2007 so that we can compare the properties of a bullish market and a bearish market before and after the latest Great Crash \cite{Han-Xie-Xiong-Zhang-Zhou-2017-FNL}. We identify the triadic time series motifs in moving windows with the length being 400 trading days. For each time series before or after 31 December 2007, we put all the $f_i$ values together and calculate the occurrence frequency distribution $P(f_i)$.

The resulting distributions are illustrated in fig.~\ref{Fig:TSM:Triadic:DJIA}. It shows that the two time series have remarkably different motif occurrence frequency distributions. Especially, the two distributions of $f_3$ are completely separated. It indicates that different markets states reflected in stock market return fluctuations can be identified by the occurrence frequencies of triadic time series motifs and there are nonlinear correlations in stock market returns that can not be modelled by geometric Brownian motions.

The concrete features of the distributions can also be understood by considering the shapes of the triadic motifs and the market trend. Since motif $M_1=(1,2,3)$ has a strictly upward trend, it appears more frequently in a bullish market than in a bearish market. As a result, $P(f_1)$ for the time series before 31 December 2007 locates on the right in fig.~\ref{Fig:TSM:Triadic:DJIA}(a). The relative locations of the two $P(f_6)$ distributions in fig.~\ref{Fig:TSM:Triadic:DJIA}(f) can be explained due to the fact that the strictly downward motif $M_6=(3,2,1)$ occurs more often in a bearish market. Similar arguments apply to $P(f_3)$ in fig.~\ref{Fig:TSM:Triadic:DJIA}(c) for the overall upward motif $M_3(2,1,3)$ and $P(f_5)$ in fig.~\ref{Fig:TSM:Triadic:DJIA}(e) for the overall downward motif $M_5(3,1,2)$.

\section{Conclusion}

In this Letter, we have introduced the concept of time series motifs for time series analysis. The occurrence frequencies of the six triadic time series motifs are derived for uncorrelated time series, which increase approximately linearly with the length of the time series. Hence, the motif profile ${\bf{f}}=(f_1, f_2, f_3, f_4, f_5, f_6)$ converges to a constant vector ${\bf{f}}=(0.2,0.2,0.1,0.2,0.1,0.2)$. These analytical results have been verified by numerical simulations. For fractional Gaussian noises, numerical simulations unveil the dependence of $f_i$ on the Hurst exponent.

We apply the time series motif analysis to the human inter-heartbeat intervals and the US stock market's DJIA index returns. We find that the certain motif occurrence frequency distributions are able to capture the different dynamics in the heartbeat rates of healthy subjects, CHF subjects and AF subjects and in the price fluctuations of bullish and bearish markets.

The time series motif analysis of real complex systems suggests the potential power of the method to distinguish different types of time series with different underlying dynamics. It is possible that the profile of the triadic time series motifs can be used to design an approach that can identify individual time series into different groups or recognize different evolutionary states of a time series. Moreover, due to the temporal information embedded in the time series motifs, the motif profile can also be used to investigate the time irreversibility of time series.

\acknowledgments

This work was supported by National Natural Science Foundation of China (11505063, 71532009) and Fundamental Research Funds for the Central Universities (222201818006).


\begin{thebibliography}{10}
\expandafter\ifx\csname url\endcsname\relax\def\url#1{\texttt{#1}}\fi

\bibitem{Brockwell-Davis-1991}
\Name{Brockwell P.~J. \and Davis R.~A.} \Book{{Time Series: Theory and Method}}
  (Springer-Verlag, New York) 1991.

\bibitem{Mandelbrot-1970-Em}
\Name{Mandelbrot B.~B.} \REVIEW{Econometrica}{38}{1970}{122}.
\newline\url{http://www.jstor.org/stable/1911596}

\bibitem{Dickey-Fuller-1981-Em}
\Name{Dickey D.~A. \and Fuller W.~A.} \REVIEW{Econometrica}{49}{1981}{1057}.

\bibitem{Bouchaud-Muzy-2003-LNP}
\Name{Bouchaud J.-P. \and Muzy J.-F.} \REVIEW{Lect. Note.
  Phys.}{636}{2003}{229}.

\bibitem{Bouchaud-Potters-Meyer-2000-EPJB}
\Name{Bouchaud J.-P., Potters M. \and Meyer M.} \REVIEW{Eur. Phys. J.
  B}{13}{2000}{595}.

\bibitem{Krawiecki-Holyst-Helbing-2002-PRL}
\Name{Krawiecki A., Holyst J.~A. \and Helbing D.} \REVIEW{Phys. Rev.
  Lett.}{89}{2002}{158701}.

\bibitem{Li-Yang-Komatsuzak-2008-PNAS}
\Name{Li C.-B., Yang H. \and Komatsuzaki T.} \REVIEW{Proc. Natl. Acad. Sci.
  U.S.A.}{105}{2008}{536}.

\bibitem{Peng-Havlin-Stanley-Goldberger-1995-Chaos}
\Name{Peng C.-K., Havlin S., Stanley H.~E. \and Goldberger A.~L.}
  \REVIEW{Chaos}{5}{1995}{82}.

\bibitem{Albert-Barabasi-2002-RMP}
\Name{Albert R. \and Barab{\'a}si A.-L.} \REVIEW{Rev. Mod.
  Phys.}{74}{2002}{47}.

\bibitem{Dorogovtsev-Mendes-2002-AdvPhys}
\Name{Dorogovtsev S.~N. \and Mendes J. F.~F.} \REVIEW{Adv.
  Phys.}{51}{2002}{1079}.

\bibitem{Newman-2003-SIAMR}
\Name{Newman M. E.~J.} \REVIEW{SIAM Rev.}{45}{2003}{167}.

\bibitem{Boccaletti-Latora-Moreno-Chavez-Hwang-2006-PR}
\Name{Boccaletti S., Latora V., Moreno Y., Chavez M. \and Hwang D.-U.}
  \REVIEW{Phys. Rep.}{424}{2006}{175}.

\bibitem{Costa-Oliveira-Travieso-Rodrigues-Boas-Antiqueira-Viana-Rocha-2011-AdvPhys}
\Name{Costa L.~F., Oliveira~Jr O.~N., Travieso G., Rodrigues F.~A., Boas P.
  R.~V., Antiqueira L., Viana M.~P. \and Rocha L. E.~C.} \REVIEW{Adv.
  Phys.}{60}{2011}{329}.

\bibitem{Barthelemy-2011-PR}
\Name{Barth{\'{e}}lemy M.} \REVIEW{Phys. Rep.}{313}{2011}{1}.

\bibitem{Holme-Saramaki-2012-PR}
\Name{Holme P. \and Saram{\"a}ki J.} \REVIEW{Phys. Rep.}{519}{2012}{97}.

\bibitem{Boccaletti-Bianconi-Criado-delGenio-GomezGardenes-Romance-SendinaNadal-Wang-Zanin-2014-PR}
\Name{Boccaletti S., Bianconi G., Criado R., del Genio C.~I., Gomez-Gardenes
  J., Romance M., Sendina-Nadal I., Wang Z. \and Zanin M.} \REVIEW{Phys.
  Rep.}{544}{2014}{1}.

\bibitem{Gao-Small-Kurths-2016-EPL}
\Name{Gao Z.-K., Small M. \and Kurths J.} \REVIEW{EPL (Europhys.
  Lett.)}{116}{2016}{50001}.

\bibitem{Zhang-Small-2006-PRL}
\Name{Zhang J. \and Small M.} \REVIEW{Phys. Rev. Lett.}{96}{2006}{238701}.

\bibitem{Zhang-Sun-Luo-Zhang-Nakamura-Small-2008-PD}
\Name{Zhang J., Sun J.-F., Luo X.-D., Zhang K., Nakamura T. \and Small M.}
  \REVIEW{Physica D}{237}{2008}{2856}.

\bibitem{Xu-Zhang-Small-2008-PNAS}
\Name{Xu X.-K., Zhang J. \and Small M.} \REVIEW{Proc. Natl. Acad. Sci.
  U.S.A.}{105}{2008}{19601}.

\bibitem{Li-Wang-2006-CSB}
\Name{Li P. \and Wang B.-H.} \REVIEW{Chin. Sci. Bull.}{51}{2006}{624}.

\bibitem{Li-Wang-2007-PA}
\Name{Li P. \and Wang B.-H.} \REVIEW{Physica A}{378}{2007}{519}.

\bibitem{Zou-Pazo-Romano-Thiel-Kurths-2007-PRE}
\Name{Zou Y., Pazo D., Romano M.~C., Thiel M. \and Kurths J.} \REVIEW{Phys.
  Rev. E}{76}{2007}{016210}.

\bibitem{Marwan-Romano-Thiel-Kurths-2007-PR}
\Name{Marwan N., Romano M.~C., Thiel M. \and Kurths J.} \REVIEW{Phys.
  Rep.}{438}{2007}{237}.

\bibitem{Marwan-2008-EPJST}
\Name{Marwan N.} \REVIEW{Eur. Phys. J.-Spec. Top.}{164}{2008}{3}.

\bibitem{Donner-Zou-Donges-Marwan-Kurths-2010-NJP}
\Name{Donner R.~V., Zou Y., Donges J.~F., Marwan N. \and Kurths J.} \REVIEW{New
  J. Phys.}{12}{2010}{033025}.

\bibitem{Marwan-Donges-Zou-Donner-Kurths-2009-PLA}
\Name{Marwan N., Donges J.~F., Zou Y., Donner R.~V. \and Kurths J.}
  \REVIEW{Phys. Lett. A}{373}{2009}{4246}.

\bibitem{Lacasa-Luque-Ballesteros-Luque-Nuno-2008-PNAS}
\Name{Lacasa L., Luque B., Ballesteros F., Luque J. \and Nu{\~n}o J.~C.}
  \REVIEW{Proc. Natl. Acad. Sci. U.S.A.}{105}{2008}{4972}.

\bibitem{Luque-Lacasa-Ballesteros-Luque-2009-PRE}
\Name{Luque B., Lacasa L., Ballesteros F. \and Luque J.} \REVIEW{Phys. Rev.
  E}{80}{2009}{046103}.

\bibitem{Gao-Cai-Yang-Dang-Zhang-2016-SR}
\Name{Gao Z.-K., Cai Q., Yang Y.-X., Dang W.-D. \and Zhang S.-S.} \REVIEW{Sci.
  Rep.}{6}{2016}{35622}.

\bibitem{Gao-Cai-Yang-Dang-2017-PA}
\Name{Gao Z.-K., Cai Q., Yang Y.-X. \and Dang W.-D.} \REVIEW{Physica
  A}{476}{2017}{43}.

\bibitem{Bianchi-Livi-Alippi-Jenssen-2017-SR}
\Name{Bianchi F.~M., Livi L., Alippi C. \and Jenssen R.} \REVIEW{Sci.
  Rep.}{7}{2017}{44037}.

\bibitem{Yang-Yang-2008-PA}
\Name{Yang Y. \and Yang H.-J.} \REVIEW{Physica A}{387}{2008}{1381}.

\bibitem{Shirazi-Jafari-Davoudi-Peinke-Tabar-Sahimi-2009-JSM}
\Name{Shirazi A.~H., Jafari G.~R., Davoudi J., Peinke J., Tabar M. R.~R. \and
  Sahimi M.} \REVIEW{J. Stat. Mech.-Theory Exp.}{}{2009}{P07046}.

\bibitem{Kostakos-2009-PA}
\Name{Kostakos V.} \REVIEW{Physica A}{388}{2009}{1007}.

\bibitem{Lacasa-Toral-2010-PRE}
\Name{Lacasa L. \and Toral R.} \REVIEW{Phys. Rev. E}{82}{2010}{036120}.

\bibitem{Ni-Jiang-Zhou-2009-PLA}
\Name{Ni X.-H., Jiang Z.-Q. \and Zhou W.-X.} \REVIEW{Phys. Lett.
  A}{373}{2009}{3822}.

\bibitem{Qian-Jiang-Zhou-2010-JPA}
\Name{Qian M.-C., Jiang Z.-Q. \and Zhou W.-X.} \REVIEW{J. Phys.
  A}{43}{2010}{335002}.

\bibitem{Yang-Wang-Yang-Mang-2009-PA}
\Name{Yang Y., Wang J.-B., Yang H.-J. \and Mang J.-S.} \REVIEW{Physica
  A}{388}{2009}{4431}.

\bibitem{Vamvakaris-Pantelous-Zuev-2018-PA}
\Name{Vamvakaris M.~D., Pantelous A.~A. \and Zuev K.~M.} \REVIEW{Physica
  A}{497}{2018}{41}.

\bibitem{Li-Zhao-2018-EPL}
\Name{Li W.-D. \and Zhao X.-J.} \REVIEW{EPL (Europhys.
  Lett.)}{122}{2018}{40007}.

\bibitem{Lacasa-Luque-Luque-Nuno-2009-EPL}
\Name{Lacasa L., Luque B., Luque J. \and Nu{\~n}o J.~C.} \REVIEW{EPL (Europhys.
  Lett.)}{86}{2009}{30001}.

\bibitem{Shao-2010-APL}
\Name{Shao Z.-G.} \REVIEW{Appl. Phys. Lett.}{96}{2010}{073703}.

\bibitem{Dong-Li-2010-APL}
\Name{Dong Z. \and Li X.} \REVIEW{Appl. Phys. Lett.}{96}{2010}{266101}.

\bibitem{Ahmadlou-Adeli-Adeli-2010-JNT}
\Name{Ahmadlou M., Adeli H. \and Adeli A.} \REVIEW{J. Neural
  Transm.}{117}{2010}{1099}.

\bibitem{Elsner-Jagger-Fogarty-2009-GRL}
\Name{Elsner J.~B., Jagger T.~H. \and Fogarty E.~A.} \REVIEW{Geophys. Res.
  Lett.}{36}{2009}{L16702}.

\bibitem{Tang-Liu-Liu-2010-MPLB}
\Name{Tang Q., Liu J. \and Liu H.-L.} \REVIEW{Mod. Phys. Lett.
  B}{24}{2010}{1541}.

\bibitem{Liu-Zhou-Yuan-2010-PA}
\Name{Liu C., Zhou W.-X. \and Yuan W.-K.} \REVIEW{Physica A}{389}{2010}{2675}.

\bibitem{Xie-Zhou-2011-PA}
\Name{Xie W.-J. \and Zhou W.-X.} \REVIEW{Physica A}{390}{2011}{3592}.

\bibitem{Ahadpour-Sadra-2012-IS}
\Name{Ahadpour S. \and Sadra Y.} \REVIEW{Inf. Sci.}{197}{2012}{161}.

\bibitem{Fan-Guo-Zha-2012-PA}
\Name{Fan C., Guo J.-L. \and Zha Y.-L.} \REVIEW{Physica A}{391}{2012}{6617}.

\bibitem{Lacasa-2014-NL}
\Name{Lacasa L.} \REVIEW{Nonlinearity}{27}{2014}{2063}.

\bibitem{Lacasa-2016-JPA}
\Name{Lacasa L.} \REVIEW{J. Phys. A}{49}{2016}{35LT01}.

\bibitem{Xie-Han-Jiang-Wei-Zhou-2017-EPL}
\Name{Xie W.-J., Han R.-Q., Jiang Z.-Q., Wei L.-J. \and Zhou W.-X.} \REVIEW{EPL
  (Europhys. Lett.)}{119}{2017}{48008}.

\bibitem{Milo-Itzkovitz-Kashtan-Levitt-ShenOrr-Ayzenshtat-Sheffer-Alon-2004-Science}
\Name{Milo R., Itzkovitz S., Kashtan N., Levitt R., Shen-Orr S., Ayzenshtat I.,
  Sheffer M. \and Alon U.} \REVIEW{Science}{303}{2004}{1538}.

\bibitem{Milo-ShenOrr-Itzkovitz-Kashtan-Chklovskii-Alon-2002-Science}
\Name{Milo R., Shen-Orr S., Itzkovitz S., Kashtan N., Chklovskii D. \and Alon
  U.} \REVIEW{Science}{298}{2002}{824}.

\bibitem{Milo-Kashtan-Itzkovitz-Newman-Alon-2004-XXX}
\Name{Milo R., Kashtan N., Itzkovitz S., Newman M. E.~J. \and Alon U.}
  \Book{{Uniform generation of random graphs with arbitrary degree sequences}}
  http://arxiv.org/abs/cond-mat/0312028 (2004).

\bibitem{Kovanen-Kaski-Kertesz-Saramaki-2013-PNAS}
\Name{Kovanen L., Kaski K., Kert{\'{e}}sz J. \and Saram{\"{a}}ki J.}
  \REVIEW{Proc. Natl. Acad. Sci. U.S.A.}{110}{2013}{18070}.

\bibitem{Klimek-Thurner-2013-NJP}
\Name{Klimek P. \and Thurner S.} \REVIEW{New J. Phys.}{15}{2013}{063008}.

\bibitem{Iacovacci-Lacasa-2016a-PRE}
\Name{Iacovacci J. \and Lacasa L.} \REVIEW{Phys. Rev. E}{93}{2016}{042309}.

\bibitem{Iacovacci-Lacasa-2016b-PRE}
\Name{Iacovacci J. \and Lacasa L.} \REVIEW{Phys. Rev. E}{94}{2016}{052309}.

\bibitem{Keller-Sinn-2005-PA}
\Name{Keller K. \and Sinn M.} \REVIEW{Physica A}{356}{2005}{114}.

\bibitem{Lacasa-Nunez-Roldan-Parrondo-Luque-2012-EPJB}
\Name{Lacasa L., Nu{\~n}ez A., Rold{\'a}n {\'E}., Parrondo J. M.~R. \and Luque
  B.} \REVIEW{Eur. Phys. J. B}{85}{2012}{217}.

\bibitem{AlvarezRamirez-Rodriguez-Echeverria-2009-Chaos}
\Name{Alvarez-Ramirez J., Rodriguez E. \and Echeverr{\'i}a J.~C.}
  \REVIEW{Chaos}{19}{2009}{028502}.

\bibitem{Aronis-Berger-Calkins-Chrispin-Marine-Spragg-Tao-Tandri-Ashikaga-2018-Chaos}
\Name{Aronis K.~N., Berger R.~D., Calkins H., Chrispin J., Marine J.~E., Spragg
  D.~D., Tao S., Tandri H. \and Ashikaga H.} \REVIEW{Chaos}{28}{2018}{063130}.

\bibitem{Han-Xie-Xiong-Zhang-Zhou-2017-FNL}
\Name{Han R.-Q., Xie W.-J., Xiong X., Zhang W. \and Zhou W.-X.} \REVIEW{Fluct.
  Noise Lett.}{16}{2017}{1750018}.

\end{thebibliography}

\end{document}